# Superconducting properties of (Ba-K)Fe$_2$As$_2$ single crystals disordered with fast neutron irradiation.

A. E. Karkin[1], T. Wolf[2] and B. N. Goshchitskii[1]

[1]Institute of Metal Physics UB RAS, 18 S. Kovalevskoi Str., 620219 Ekaterinburg, Russia
[2]Institut für Festkörperphysik, Karlsruhe Institute of Technology, 76021 Karlsruhe, Germany

e-mail: aekarkin@rambler.ru

Resistivity $\rho(T)$, Hall coefficient $R_H(T)$, superconducting transition temperature $T_c$ and slopes of the upper critical field $dH_{c2}/dT$ were studied in (Ba$_{1-x}$K$_x$)Fe$_2$As$_2$ ($x$ = 0.218, 0.356, 0.531) single crystals irradiated with fast neutrons. It is found that $dT_c/d\rho_{SC}$ - the rate of decreasing $T_c$ as a function of the $\rho_{SC}$ ($\rho_{SC}$ is the resistivity at $T = T_c$) - linearly increases with concentration $x$. Slow changes in the Hall coefficient $R_H$, as well as the quadratic electronic contribution to the resistivity, show that there are no substantial changes in the topology of the Fermi surface caused by irradiation. The slopes of the upper critical field $dH_{c2}/dT$ in $ab$- and $c$-directions as a function of $\rho_{SC}$ determined by Hall measurements show a reasonable agreement with a model that suggests constancy of the band parameters.

Soon after the discovery of superconductivity in 2008 in the Fe-based compound LaFeAs(O,F) with the transition temperature $T_c$ = 26 K [1], the effect of neutron irradiation on superconducting properties of this compound was studied [2]. Radiation defects that are nonmagnetic centers of scattering electrons were shown to cause fast depression of superconductivity, similarly to the case of Cu-based superconductors [3, 4]. Analogous effects were also observed in the systems NdFeAs(O,F) upon irradiation with $\alpha$-particles [5] and in Ba(Fe$_{1-x}$Co$_x$)$_2$As$_2$ upon irradiation with 3 MeV protons [6]. Suppression of superconductivity by defects or nonmagnetic impurities is an indication of unconventional (non-phononic) superconductivity in systems with a sign-reversible gap function [7, 8, 9]. On the contrary, upon irradiation of systems with the electron-phonon mechanism of superconductivity, such as Nb$_3$Sn [10] or MgB$_2$ [11, 4], the superconductivity is preserved despite the fact that $T_c$ also falls because of the decrease in the density of states at the Fermi level $N(E_F)$.

We studied the effects of irradiation with fast neutrons with a fluence of $5*10^{18}$ cm$^{-2}$ in the (Ba$_{1-x}$K$_x$)Fe$_2$As$_2$ system at the temperature of irradiation $T_{irr}$ = 330±10 K. Recently, results of alike studies of this system subjected to irradiation with 3 MeV protons have been reported [12]. In this work, the main attention was paid to the determination of upper critical fields in the $ab$- and $c$-directions via measuring resistivity $\rho(T)$ and Hall coefficient $R_H(T)$ in initial and irradiated samples.

Since substitution of K for Ba is unlikely to create a noticeable disorder in the FeAs plane [13], one can expect a low concentration of proper defects in the initial state, which makes it possible, in particular, to observe the transition from the clean limit (mean free path $l$ is larger than the coherence length $\xi$) to the dirty limit ($l < \xi$). Compositions were chosen with $x$ = 0.218, 0.356, and 0.531, which can be assigned to underdoped, optimally doped, and overdoped systems, respectively.

The existing data on the values of slopes, $dH_{c2}/dT$, near $T_c$ in Fe-based systems are strongly dependent on the method of measurement. The uncertainty arisen is caused by the presence of superconducting fluctuations in a large temperature range above $T_c$ that give a remarkable contribution to conductivity, as well as spatial compositional inhomogeneities on a scale comparable to the coherence length (on the order of tens Å); both effects result in a noticeable broadening of the superconducting transition. Thus, the resistivity method of determination of $dH_{c2}/dT$, for example, gives different results depending on a criterion chosen for the determination of $T_c(H)$ value and accuracy of separation of the fluctuation contribution. Likely, a more substantiated method in terms of physics is the method of measuring electron heat capacity in which the determination of $T_c(H)$ value is based on equality of entropies in the normal and superconducting states. Another appropriate method is the analysis of temperature dependences of nondiagonal components of resistivity near $T_c$. This method, as will be shown in what follows, gives more realistic values of $dH_{c2}/dT$ due to the absence of a fluctuation contribution to the Hall coefficient at the orientation $H \parallel c$ and, besides, the absence at $T > T_c$ of the normal contribution at $H \parallel ab$.

Single crystals of (Ba, K)Fe$_2$As$_2$ were grown in alumina crucibles using a self flux method with ratios (Ba, K) : FeAs = 1:4.5 - 1:5 [14]. The crucibles were sealed in an iron cylinder filled with argon gas. After heating up to 1150-1160°C the furnace was cooled down slowly at rates between 0.3-0.41°C/h. Near 1040°C the furnace was turned over to separate remaining liquid flux from the grown crystals and then switched off.

Measurements of resistivity by the Montgomery method [15, 16] on samples with characteristic dimensions in the $ab$-plane 1×0.5 mm$^2$ and a thickness of 50-100 μm showed that the values of in-plane resistivity $\rho_{ab}$ weakly depend on composition; the absolute magnitudes of $\rho_{ab}$ at room temperature are on the order of 300 μΩcм, which agrees well with the data of [17, 18, 12, 19]. On the contrary, the out-of-plane resistivity $\rho_c$ significantly varies depending both on composition and from sample to sample, so that the anisotropy of resistivity $\rho_c/\rho_{ab}$ changes within the limits from 5 to 50. As will be shown below, based on the analysis of the anisotropy of upper critical field, the expected value of $\rho_c/\rho_{ab}$ characterizing the resistivity anisotropy should be on the order of 2, which is significantly less than experimental data. Indeed, cracks and flux inclusions usually are oriented within $ab$-plane which influences $\rho_c$ much more than $\rho_{ab}$. Thus, there are enough grounds to not consider $\rho_c$ as an intrinsic property; therefore, herein we will not consider the behavior of $\rho_c$ upon irradiation.

The in-plane values of $\rho(T)$ in the normal state, shown for two limiting compositions with $x$ = 0.218 and 0.531 in Fig. 1, demonstrate a typical metallic behavior upon irradiation: virtually parallel shift along the ordinate axis (Mattissen rule). Note a stronger increase of the residual resistivity $\rho_0$ for $x$ = 0.531 in comparison with $x$ = 0.218.

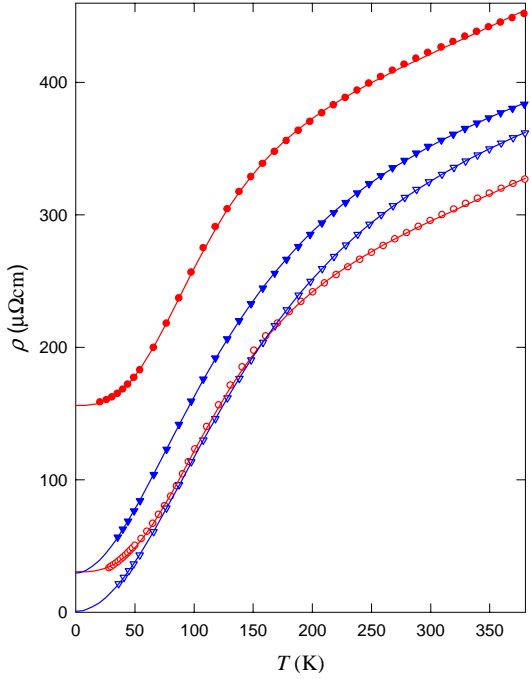

Fig. 1. Temperature ($T$) dependences of in-plane resistivity $\rho$ for initial (open symbols) and irradiated (closed symbols) samples (Ba$_{1-x}$K$_x$)Fe$_2$As$_2$, $x = 0.218$ (red circles) and 0.531 (blue triangles).

To describe the saturation effects for $\rho(T)$ at high temperatures, the so-called shunting model is often used:

$$1/\rho = 1/\rho_1 + 1/\rho_2, \quad (1)$$

which employs two conductivity channels. The first has a typical form for metals:

$$\rho_1 = A_0 + A_n T^n, \quad (2)$$

where $n = 2$ for the predominant electron-electron scattering, $\rho_1 = A_0 + A_2 T^2$. The existence of the second channel ($\rho_2$) can be treated differently. It can be considered as a non-coherent conductivity channel [20] for which $\rho_2$ should poorly depend on temperature; such interpretation seems to be supported by optical measurements [21]. The second channel can be also related to the presence of a small group of charge carriers with the effective mass much smaller that that of the main carriers and, consequently, weaker temperature dependence. Then

$$\rho_1 = B_0 + B_2 T^2, \quad (3),$$

with conditions $A_0 \ll B_0$, $A_2 \gg B_2$. However, it is not clear whether the carriers with such properties do exist. Furthermore, the observed effect of $\rho$ saturation can be a sequence of the Ioffe-Regel rule: electron scattering becomes ineffective when the free-path length for electrons $l$ is less than the reciprocal wave vector $2\pi k_F$. In this case, in expression for electrical conductivity $\sigma \sim (k_F)^2 l$, instead of $l$ there should be a value of the order of the lattice parameter $a_0 \approx (2\pi k_F)^{-1}$. The interpolation formula $\sigma \approx (k_F)^2 l + k_F/(2\pi)$ is virtually equivalent to Eq. 1.

Fitting of the experimental curves $\rho(T)$ showed that $n = 2.85 \pm 0.03$ for $x = 0.218$, $2.15 \pm 0.03$ for $x = 0.356$, and $1.95 \pm 0.05$ for $x = 0.531$, and does not change upon irradiation. For the two last compositions the value of $n$ is close to 2, which corresponds to predominant electron-electron scattering. For the first composition, $n$ is near 3, which, however, can be explained by the closeness of superconducting transition to structural (magnetic) transitions at $T_s \approx 80$ K in the initial sample [22, 23] and $T_s \approx 50$ K in the irradiated sample, which are clearly seen in the temperature dependences of the Hall coefficient $R_H(T)$, Fig. 3.

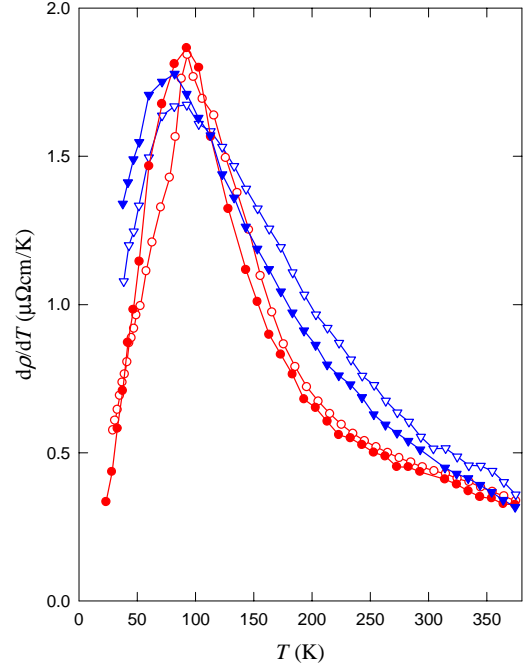

Fig. 2. Temperature dependences of derivative of the in-plane resistivity $d\rho/dT$ of (Ba$_{1-x}$K$_x$)Fe$_2$As$_2$ samples; curve notations are the same as in Fig. 1.

Yet, the derivative plots $d\rho/dT$ (Fig. 2) clearly show that the slope (i.e., coefficient $a_2$ proportional to $\{N(E_F)\}^2$) is approximately the same for different compositions and virtually does not change upon irradiation. Since changes in $R_H(T)$ also are insignificant (Fig. 3), one can conclude on smallness of changes in the band structure upon irradiation of (Ba$_{1-x}$K$_x$)Fe$_2$As$_2$ with a fluence of $5*10^{18}$ cm$^{-2}$.

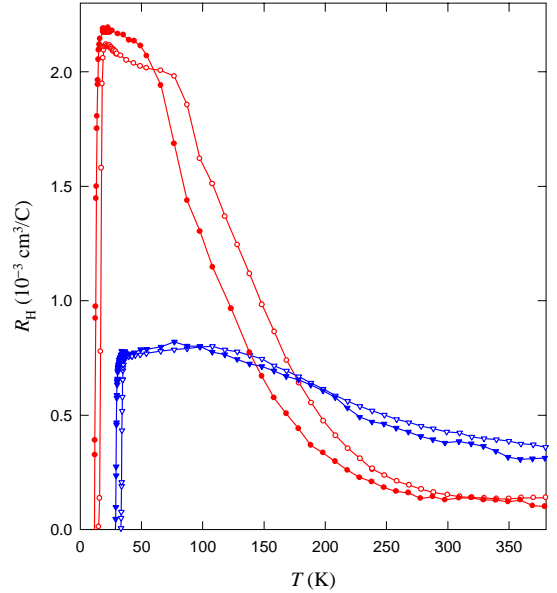

Fig. 3. Temperature dependences of the in-plane Hall coefficient $R_H$ in a magnetic field $H = 13.0$ T for samples (Ba$_{1-x}$K$_x$)Fe$_2$As$_2$; curve notations are the same as in Fig. 1.

In Fig. 4 the values of $T_c$ are shown as a function of $\rho_{SC} = \rho(T = T_c)$ for the initial and irradiated samples of (Ba$_{1-x}$K$_x$)Fe$_2$As$_2$ ($x = 0.218, 0.356$, and $0.531$) together with the data on proton irradiation of the compounds Ba(Fe$_{1-x}$Co$_x$)$_2$As$_2$ ($x = 0.045, 0.075$, and $0.113$) [6] and (Ba$_{1-x}$K$_x$)Fe$_2$As$_2$ ($x = 0.23, 0.42$, and $0.69$) [12]; by composition, the two last compounds can be ascribed to underdoped, optimally doped, and overdoped systems, respec-tively. Here, instead of the residual resistivity $\rho_0$ (or increment in the electrical resistivity, $\Delta\rho$, as it was done, for example, in [6, 12]), we use $\rho_{SC} = \rho(T = T_c)$ to take into account all possible kinds of nonmagnetic scattering: at defects, phonons, and electrons.

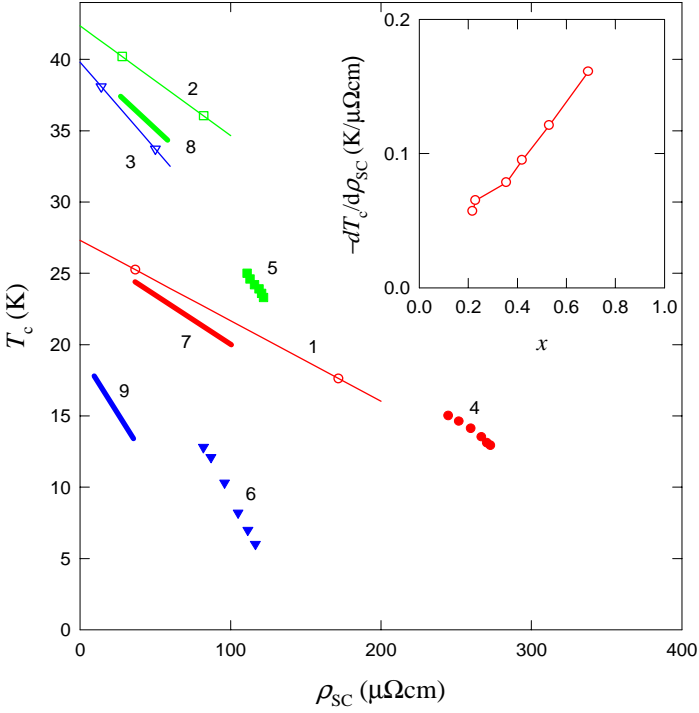

Fig. 4. $T_c$ as a function of $\rho_{SC}$ for initial and irradiated samples of $(Ba_{1-x}K_x)Fe_2As_2$ from our measurements (open symbols) for $x = 0.218$ (red circles), 0.356 (green squares) and 0.531 (blue triangles). Data on proton irradiation are also presented for compounds $Ba(Fe_{1-x}Co_x)_2As_2$ (closed symbols): for $x = 0.045$ (red circles), 0.075 (green squares) and 0.113 (blue triangles) [6] and $(Ba_{1-x}K_x)Fe_2As_2$ (red bold line), $x = 0.23$ (green bold line), 0.42 and 0.69 (blue bold line) [12]. In inset, the slope $-dT_c/d\rho_{SC}$ is shown as a function of $x$ in $(Ba_{1-x}K_x)Fe_2As_2$.

Note a very similar behavior of $T_c$ depending on $\rho_{SC}$ in the system $(Ba_{1-x}K_x)Fe_2As_2$ upon neutron and proton types of irradiation for compositions with similar $x$. Besides, for the system $Ba(Fe_{1-x}Co_x)_2As_2$ the rate of decreasing the superconducting transition temperature, $dT_c/d\rho$, turns out nearly equal for samples with corresponding levels of doping (underdoped, optimally doped and overdoped). Figure 4 clearly shows that lower $T_c$ values in the system $Ba(Fe_{1-x}Co_x)_2As_2$ in comparison with $(Ba_{1-x}K_x)Fe_2As_2$ are traceable to a large initial disorder.

To determine values of the upper critical field near $T_c$, we used samples of characteristic dimensions in the $ab$-plane of $1.5\times1.5$ mm$^2$ and a thickness of 20-40 μm with symmetrical current and potential contacts. We measured the values of resistivity $r_{\alpha\beta\gamma} = U_\alpha/I_\beta$, where the two first indices $\alpha$ and $\beta$ are two mutually perpendicular directions in the $ab$-plane that correspond to positions of potential and current contacts, respectively, and the third index $\gamma$ stands for the direction of magnetic field $H$. The reverse of current and potential contacts leads to a change in the sign of measured voltage $U_\alpha$ (as in the case of changing the polarity of $H$), therefore, the Hall resistivity $\rho_{\alpha\beta\gamma} = d(r_{\alpha\beta\gamma} - r_{\beta\alpha\gamma})/2$, where $d$ is the sample thickness. Correspondingly, the Hall constant (shown in Fig.3) will be determined as $R_H = \rho_{abc}/H$, $\rho_{aba} = 0$ at $T > T_c$, whereas $\rho_{aaa}$ and $\rho_{aac}$ are the resistivities in the directions parallel and perpendicular to the plane $ab$.

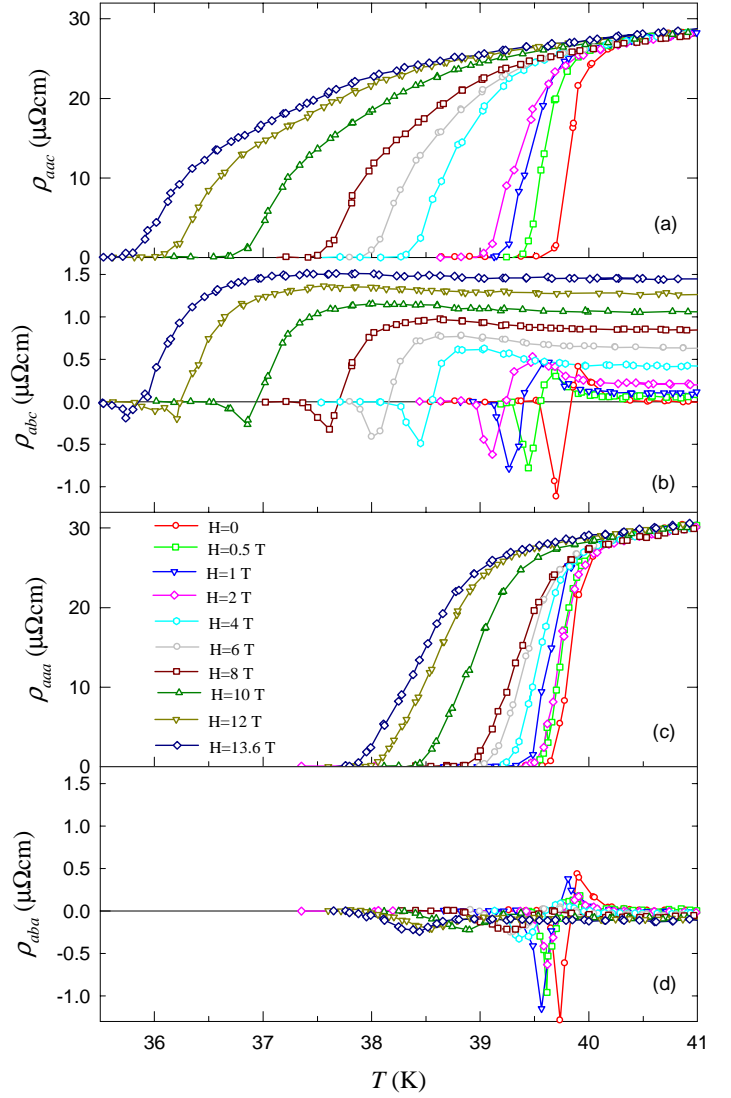

Fig. 5. Resistivities $\rho_{\alpha\beta\gamma}$ curves for the initial sample of $(Ba_{1-x}K_x)Fe_2As_2$ ($x = 0.356$) close to the superconducting transition; $\alpha$ and $\beta$ are two mutually perpendicular directions in the plane $ab$ corresponding to the positions of potential and current contacts, respectively; $\gamma$ is the direction of magnetic field $H$.

Characteristic temperature dependences of $\rho_{\alpha\beta\gamma}$ are shown in Fig.5 for the initial sample of $(Ba_{1-x}K_x)Fe_2As_2$ ($x = 0.356$) (here, a weak negative signal in $\rho_{aba}$ is connected with a small, on the order of 3°, disorientation of $H$ with respect to the plane $ab$). In the temperature dependencies of resistivities $\rho_{aaa}$ and $\rho_{aac}$, especially in large fields, it is rather difficult to determine a characteristic point that would match the critical temperature $T_c(H)$. This is mainly related to the presence of a fluctuation region which, according to the direct measurements of the diamagnetic response [24], is by 2-5 K higher than $T_c$. On the contrary, in the curves of the Hall resistivities $\rho_{aba}$ and $\rho_{abc}$ in the superconducting state there are observed two well pronounced extremes, minimum and maximum. It is natural to relate the $T_c(H)$ values to the positive deviations of $\rho_{abc}$ from a constant value at $T > T_c$ and of $\rho_{aba}$, from the zero line.

Here, we will not discuss the reasons for the appearance of a signal of such complicated form in $\rho_{aba}$ and $\rho_{abc}$ at $T \leq T_c$ and its absolute absence in the fluctuation region at $T > T_c$. The $\rho_{aba}(T)$ and $\rho_{abc}(T)$ curves do not depend on the magnetic field directions and they are reversible (ZF and ZFC curves are the same).

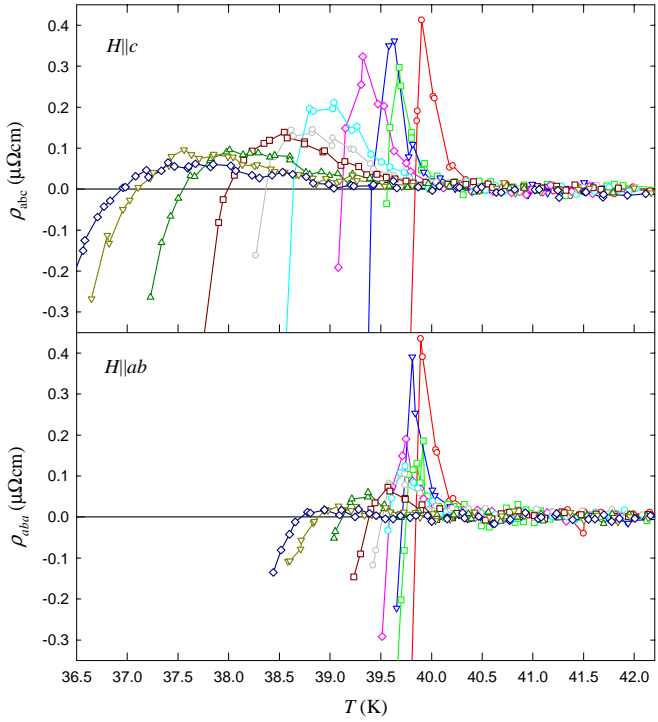

Fig. 6. The same $\rho_{\alpha\beta\gamma}$ curves as in Figs. 5b, 5d on a large scale (the constant contribution at $T > T_c$ is subtracted).

In Fig. 6 curves $\rho_{aba}(T)$ and $\rho_{abc}(T)$ are shown on a larger scale, the constant contribution at $T > T_c$ being subtracted. Although the positive deviation from the zero line in large fields is not very large, this way of determination of $T_c(H)$ is evidently more correct than employment of a certain "criterion" in the resistivity curves (Fig. 5). On the whole, the "Hall" method gives remarkably higher values of $-dH_{c2}/dT$ in comparison with the "resistivity" method; yet, they are closer to the physically substantiated method of "electron heat capacity" [25, 26, 27].

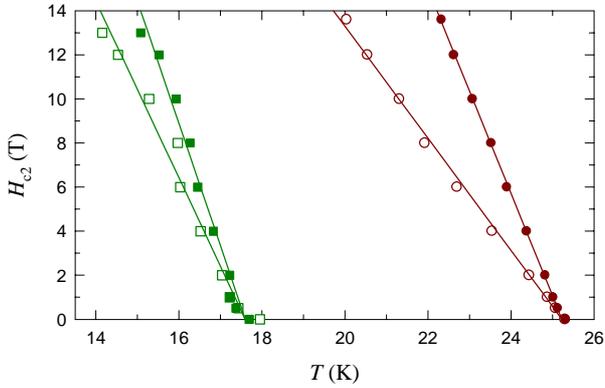

Fig. 7. Temperature dependences of the upper critical field $H_{c2}$ for initial (dark red circles) and irradiated (dark green squares) samples of $(Ba_{1-x}K_x)Fe_2As_2$, $x = 0.218$, in $ab$ (open symbols) and $c$ (closed symbols) directions.

In Fig. 7, a typical example of the temperature dependences $H_{c2}(T)$ is shown for the initial and irradiated samples of $(Ba_{1-x}K_x)Fe_2As_2$, $x = 0.218$, in $ab$ and $c$ directions. Experimental points for the initial sample fall rather well on the straight line; for the irradiated sample there is observed a remarkable deviation from the line, which is explained by the lower maxima in $\rho_{aba}(T)$ and $\rho_{abc}(T)$. This causes significant uncertainties, in particular, when analyzing dependences of the upper critical fields on the resistivity $\rho_{SC}$ (Fig. 8).

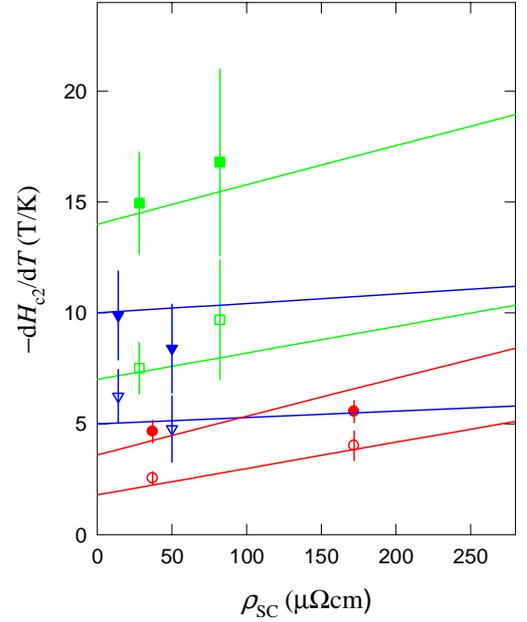

Fig. 8. The slope $dH_{c2}/dT$ as a function of $\rho_{SC}$ for $H \parallel c$ (open symbols) and $H \parallel ab$ (closed symbols) in the samples of $(Ba_{1-x}K_x)Fe_2As_2$ for $x = 0.218$ (red circles), 0.356 (green squares) and 0.531 (blue triangles). Solid lines present fitting curves, Eqs. 8, 9.

However, for a qualitative analysis, we can write the slope of the upper critical field in the form

$$-dH_{c2}/dT = \phi_0/(0.69 \cdot 2\pi \xi^2 T_c),$$

where the coherence length $\xi$ can be written in the intermediate limit as

$$1/\xi^2 \approx (1/\xi_0)(1/\xi_0 + 1/l),$$

here $\xi_0$ is related to the clean limit, i.e., at $\rho_0 \to 0$. Introducing designation $h_\alpha = -dH_{c2,\alpha}/dT$, where the field $H$ is parallel to the direction $\alpha$ ($\alpha$ denotes $ab$ or $c$), and accounting for $\xi_{0,\alpha} = (\hbar v_{F,\alpha})/(2\pi k_B T_c)$, where $v_F$ is the Fermi velocity, we obtain

$$h_c = \phi_0/(0.69 \cdot 2\pi (\xi_{ab})^2 T_c) = h_{0c}(T_c/T_{c0} + \xi_{0ab}/l), \quad (4)$$

$$h_{ab} = \phi_0/(0.69 \cdot 2\pi \xi_c \xi_{ab} T_c) = h_{0ab}\{(T_c/T_{c0} + \xi_{0,c}/l)(T_c/T_{c0} + \xi_{0ab}/l)\}^{1/2} = h_{0c}k\{(T_c/T_{c0} + \xi_{0ab}/(kl))(T_c/T_{c0} + \xi_{0ab}/l)\}^{1/2}, \quad (5)$$

where index 0 in the value $h_{0\alpha}$ again stands for the limit $\rho_0 \to 0$ and $k = h_{0ab}/h_{0c}$ is the anisotropy of the upper critical field in the limit $\rho_0 \to 0$.

To specify the interrelation of the experimental values $h_\alpha$, $T_c$, and $\rho_{SC}$, note that the dependence of $T_c$ on $\rho_{SC}$ can be extrapolated by the linear function $T_c/T_{c0} = (1 - b_1\rho_{SC})$, as is shown in Fig. 4. There are different ways how to pass from the microscopic values $\xi_{0,ab}/l$ to the experimentally observed $\rho_{SC}$, which are described in detail in [5, 6, 13]. Here, we use another approach. According to the Ioffe-Regel rule, in the limit of strong electron scattering, the mean free path $l$ should be on the order of the lattice parameter $a_0 \approx 4$ Å. Since the band parameters do not change upon irradiation, we can set

$$\xi_{0ab}/l = b_2\rho_{SC}, \quad (6),$$

where the coefficient $b_2$ is determined from expression

$$\xi_{0ab}/a_0 = b_2\rho_\infty, \quad (7),$$

which defines $\rho_\infty$ as a high-temperature limit of the resistivity by formula (1). This value falls in the range 800-1000 μΩcm; for simplicity, we take the medium value $\rho_\infty$ = 900 μΩcm for all compositions. Then, $\xi_{0ab}$ will be determined from the relationship $h_{0c} = \phi_0/(0.69 \cdot 2\pi(\xi_{0ab})^2 T_c)$, where for $h_{0c}$ there stand the corresponding values for the initial samples of each composition. Hence, the relationship (4) and (5) will acquire a simple form

$$h_c = h_{0c}\{1 + (b_2 - b_1)\rho_{SC}\}, \quad (8)$$
$$h_{ab} = h_{0c}k\{[1 + (b_2 - b_1)\rho_{SC}][1 + (b_2 - b_1)\rho_{SC}/k)]\}^{1/2}. \quad (9)$$

Thus, in this approximation ($T_c$ linearly depends on $\rho_{SC}$ and $\xi_{0ab}$ does not change upon irradiation), the rate of changing the slopes of the upper critical fields upon increasing $\rho_{SC}$ is determined by the ratio of the coefficients $b_2$ and $b_1$; the first depends on the rate of decreasing $T_c$ upon irradiation (Fig. 4), the second, depends on the values of the upper critical fields for the initial samples $h_{0c}$ and $h_{0ab}$.

In Fig. 8, the experimental dependences of the upper critical fields for the initial and irradiated samples of $(Ba_{1-x}K_x)Fe_2As_2$ ($x$ = 0.218, 0.356, and 0.531) in $ab$ and $c$ directions are shown together with the results of fitting with the use of Eqs. 8, 9; here, for simplicity, we take the value averaged over the samples of different compositions, $k$ = 2. Despite the fact that the supposition $l \sim a_0$ (Eq. 6) is rather rough, Eqs. 8, 9 reproduce rather well the tendencies for changes of the upper critical fields as a function of $\rho_{SC}$. Thus, for example, the relatively weak dependences of $h_c$ and $h_{ab}$ on $\rho_{SC}$ for x = 0.531 are traceable to a more rapid decrease of $T_c$ with $\rho_{SC}$, so that $b_2 \approx b_1$. Besides, as Fig. 8 shows, the large field values for $(Ba_{1-x}K_x)Fe_2As_2$ (and feasibly for other Fe-based superconductors) are connected with low $v_F$ rather than high values of the residual resistivities. The estimation of the $\xi_0/l$ using Eq. 6 and experimental values of $\rho_{SC}$ and $b_2$ for initial samples gives 0.64, 0.20 and 0.10 for $x$ = 0.218, 0.356 and 0.531, respectively; that corresponds to clean limit $\xi_0/l < 1$. Because in the clean limit $\rho_c/\rho_{ab} = \xi_{ab}/\xi_c = k$, and $k \approx 2$ in our estimations, we can expect $\rho_c/\rho_{ab} \approx 2$ in initial samples, as was mentioned above.

As for the dependence of $T_c$ on $\rho_{SC}$, for comparison with the theoretical models, we made use of the universal Abrikosov-Gor'kov (AG) equation describing the superconductivity suppression by magnetic impurities for the case of $s$-pairing, and by nonmagnetic impurities (defects) for the case of $d$- and $s^\pm$-pairing [28, 29, 30]:

$$\ln(1/t) = \psi(g/t + 1/2) - \psi(1/2), \quad (10)$$

where $g = \hbar/(2\pi k_B T_{c0} \tau) = \xi_0/l$, $\psi$ is the digamma function, $t = T_c/T_{c0}$, $T_{c0}$ and $T_c$ are the superconducting temperatures of the initial and disordered systems, respectively, $\tau$ is the electronic relaxation time. Equation (10) describes the decrease of $T_c$ as a function of the inverse relaxation time $\bar\tau^{-1}$; superconductivity is suppressed at $g > g_c = 0.28$. According to (6), $g$ can be written in the form

$$g = \xi_{0ab}/l = b_2\rho_{SC}. \quad (11)$$

This estimate gives $g_c$ = 0.28 at $\rho_{SC}$ = 20-40 μΩcm for $x$ = 0.218, 0.356, and 0.531, which is much lower than the experimental values $\rho_{SC}$ = 200-400 μΩcm at which superconductivity disappears (Fig. 4). Similar results follow from the estimates made in [12].

Besides, it is necessary to explain a virtually linear increase of the rate of decreasing $T_c$ with concentration $x$. A guess arises that it is connected merely with a decrease in the electron concentration $n_e$, which can be expected with increasing $x$; then from the relationship $\rho = m^*/(n_e e^2 \tau)$ at a constant effective mass $m^*$ we formally obtain that the rate $dT_c/d(1/\tau)$ slowly depends on $x$.

Such supposition, however, evidently contradicts the many-band nature of the compound $(Ba_{1-x}K_x)Fe_2As_2$ [31, 32, 33], which gives rise to a rather complicated temperature dependence of $R_H$ (Fig. 3). The relatively weak change of the value $\rho$ versus $x$ at room temperature (Fig. 3) also disagrees with the idea that electrons make the main contribution to resistivity $\rho(T)$.

In conclusion, our results show the fast decrease of the superconducting temperature $T_c$ in the $(Ba_{1-x}K_x)Fe_2As_2$ ($x$ = 0.218, 0.356, and 0.531) samples under the fast-neutron irradiation. The rate of decreasing $T_c$ as a function of $\rho_{SC}$ (resistivity at $T = T_c$) linearly grows with concentration $x$. The slow changes in the Hall coefficient $R_H$, as well as the quadratic electronic contribution to the resistivity, show that there are no substantial changes in the topology of the Fermi surface caused by irradiation. Therefore the suppression of superconductivity by defects is an indication of unconventional (non-phononic) superconductivity in this system. The slopes of the upper critical field $dH_{c2}/dT$ in the $ab$- and $c$-directions as a function of $\rho_{SC}$ determined from the Hall measurements show a reasonable agreement with the model suggesting a constancy of the band parameters (a rigid band model).

The work was carried out according to IMP Program "Potok" with a partial support of the Program for Basic Research of the Presidium of RAS "Quantum mesoscopic and disordered structures" (project № 12-P-2-1018 of the Ural Branch of RAS), Russian Foundation for Basic Research (project № 11-02-00224), and State Contract of the the Ministry of Education and Science (project № 14.518.11.7020).